\documentclass[a4paper,11pt]{article}
\usepackage[utf8]{inputenc}
\usepackage{amsmath,amsfonts}
\usepackage{graphicx}

\newcommand{\numgroups}{g}

\title{An efficient algorithm for structured sparse quantile regression}

\author{Vahid Nassiri${}^a$ and Ignace Loris${}^b$\\ ${}^a$Department of Mathematics, Vrije Universiteit Brussel, Belgium,\\ ${}^b$Department of Mathematics, Université Libre de Bruxelles, Belgium}

\begin{document}

\maketitle

\begin{abstract}
Quantile regression is studied in combination with a penalty which
promotes structured (or group) sparsity. A mixed $\ell_{1,\infty}$-norm on the parameter vector is used to impose structured sparsity on the traditional quantile regression problem. An algorithm is derived to calculate the piece-wise linear solution path of the corresponding minimization
problem. A Matlab implementation of the proposed algorithm is provided and some applications of the methods are also studied.

\end{abstract}

\section{Introduction}\label{introductionsection}

As \cite{Gabaix2008} have remarked, a good statistical model has
seven key properties: 1. Parsimony, 2. Tractability, 3. Conceptual
insightfulness, 4. Generalizability, 5. Falsifiability, 6. Empirical
consistency, and 7. Predictive precision. In this paper a structured
sparse quantile regression model is studied and an efficient algorithm is
proposed to solve the corresponding minimization problem. As an
illustration, two applications are discussed where such a model is
preferable (in the sense of several of the above properties) to
others, and where the proposed algorithm is more useful than others.

The quantile regression model of \cite{Koenker1978} allows for
studying the effect of explanatory variables on the entire
conditional distribution of the response variable, and not only on
its center. In this sense quantile regression provides a deeper
\emph{conceptual insightfulness} into data than least squares
models. Since \emph{parsimony} is a key property of a good model,
variable selection techniques have attracted a lot of attention in
recent statistical literature, \cite{Hjort2008}. One reason
that sparse
models have become popular is the availability of very large data
sets; see e.g. \cite{FanLi2006}. A promising approach to achieve a
sparse model is to penalize models with a sparsity promoting
penalty. Among different penalties, the $\ell_1$-norm is the most
popular one. It has e.g. been used in combination with least squares
in Lasso regression of \cite{Tibshirani1996}, and an effective algorithm was proposed by \cite{OsPrT2000} and \cite{EfHJT2004}. Authors such as
\cite{Li2008} and \cite{Fuchs2009} have proposed similar algorithms
for the $\ell_1$-norm penalized quantile regression problem.

Although the $\ell_1$-norm penalty does a good job in selecting individual variables,
there is no control on the groups of variables it selects. There are
cases where one is interested in selecting a group of variables
instead of individual variables. In some cases it is inevitable, see \cite{Cohen1991}.
Consider e.g. race as a predictor with three levels:
black, white, and other. The standard approach is making two dummy
variables ($X_1, X_2$) out of it: $(0,0)$ for other, $(1,0$) for
black, and $(0,1)$ for white. Obviously in such a case one factor
is represented by two variables. Therefore, one may select the
pair $(X_1, X_2)$ together, or select none of them. The usual
$\ell_1$-norm penalty cannot guarantee such behavior. Finding an alternative to
the $\ell_1$-norm seems therefore to be necessary in the cases
where groups of variables should be selected instead of individual variables.
As an individual can be considered as a singleton group, the grouping approach should be a
\emph{generalization} of the usual $\ell_1$-norm penalty.

An appropriate penalty for the grouping case would be a mixed norm
which applies $\ell_1$-norm penalty to groups, in order to promote
sparsity of groups of variables. Within each group the
$\ell_\infty$-norm (also known as the $\max$-norm) can be used to
ensure that, once a single member of a group is chosen (is nonzero),
all other members of the same group can grow to the same size
without penalty. Other choices are possible too, but we restrict
ourselves to this mixed $\ell_1-\ell_\infty$-norm (or
$\ell_{1,\infty}$-norm for short) choice. \cite{Yuan2006} have
introduced such a group sparsity penalty, and have solved the problem
for the least squares model and have applied it to a low birth weight data set.

In this paper we formulate an algorithm for
$\ell_{1,\infty}$-norm penalized quantile regression
instead of least squares regression. As \emph{tractability} is a key
property of a good model, we choose the mixed norm penalty function
because the corresponding models have a simple dependence on the
penalty parameter. Choosing the $\ell_{\infty}$-norm within groups
guarantees that the model is a piece-wise linear function both of
the penalty parameter and of the $\ell_{1,\infty}$-norm of the
model. Other choices (such as e.g. $\ell_{1,2}$) would lead
to less tractable optimization problems. The main
contribution of the present paper is the description of an efficient
computational algorithm for solving this optimization problem. This
algorithm applies only to the case of non-overlapping groups. A
Matlab implementation of the algorithm is provided on the authors'
webpage \cite{Lorisweb2012}.


In order to illustrate the performance of the proposed penalized
model and the corresponding optimization algorithm, we study two main applications: Firstly the use of qualitative explanatory variables with more than two levels, and
secondly, simultaneous variable selection for a vector of response
variables.

For the first application, we analyze a data set pertaining to `low
birth weight' (LBW). Low birth weight is a common subject in quantile
regression literature (see e.g. \cite{Koenker2001}). If one is
interested in the effect of different variables on the lower tail of
the conditional distribution of the infant's
weight, least squares models are not appropriate, because they
analyze the effect of the explanatory variables on the conditional
mean of the response variable. On the other hand, using quantile
regression, one can study lower conditional quantiles of the
response. Due to presence of a qualitative variable with more than
two levels in the model, the sparsity promoting $\ell_1$-norm
penalty is not appropriate either. Therefore, a penalized model with
a mixed norm penalty seems to be more appropriate for selecting
effective variables and estimating their effect.

For the second application, the `93CARS' data set of \cite{Amstat}
is studied. It consists of $14$ explanatory variables and $5$
response variables which are highly correlated. \cite{TURLACH2005}
have suggested that a simultaneous variable selection approach is
more interesting for modeling a vector of correlated response
variables with a common set of explanatory variables. Using the
proposed mixed norm as penalty one can put the coefficients of each
explanatory variable for all the response variables in a group and
perform a simultaneous variable selection. While \cite{TURLACH2005}
worked it out in the least-squares framework, we use quantile
regression which is more robust and makes it possible to study the
whole conditional distribution and not just its center. \cite{Zoua2008} have studied the latter problem too,
using a standard linear programme solver. In contrast, our proposed
algorithm produces the whole piece-wise linear solution
path (for different values of the penalty parameter) which makes the
variable selection computationally more efficient.

Section~\ref{problemsection} states the structured sparse quantile
regression problem and introduces notation. The main tools for
describing the corresponding minimization problem are
discussed in Section~\ref{solutionsection}. The algorithm to solve
$\ell_{1,\infty}$-norm penalized quantile regression problem is
introduced and discussed in Section~\ref{algorithmsection}.
Section~\ref{applicationsection} illustrates the use of structured
quantile regression for the analysis of a low birth weight data set.
A simultaneous variable selection problem for a vector of response
variables data set is also examined. The paper is concluded in
Section~\ref{conclusionsection}.

\section{Problem statement}
\label{problemsection}

In this paper, $X_{n\times m}=(X_1,\ldots,X_m)$ is the design matrix containing the explanatory variables.
The linear model which we study is:
\begin{equation}\label{model}
y=X\beta +\epsilon
\end{equation}
The loss function which quantile regression (see
\cite{Koenker1978}) tries to minimize is given by:
\begin{equation}\label{lossfunction}
\sum_{i=1}^{n} \varrho_{\tau}\left((y-X\beta)_i \right)
\end{equation}
where the function $\varrho_{\tau}$ is defined as:
\begin{equation}
\label{loss} \varrho_{\tau}(t) = \left\{
\begin{array}{rl}
 2\tau\, t& \text{if } t \geq 0\\
-2(1-\tau)\,t & \text{if } t\leq 0,
\end{array} \right.
\end{equation}
for $t\in\mathbb{R}$ and $0<\tau<1$. For $\tau=1/2$, one recovers
$\varrho_\tau(t)=|t|$. This loss function is more robust to outliers
than the usual quadratic one: the minimizer of the least squares loss
is the mean, while the minimizer of the least absolute deviations loss (expression (\ref{lossfunction}) with $\tau=1/2$)
is the median. The optimization problem encountered
in quantile regression thus is:
\begin{equation}\label{qrproblem}
\hat\beta_\mathrm{QR}=\arg\min_{\beta}\sum_{i=1}^{n} \varrho_{\tau}\left((y-X\beta)_i \right).
\end{equation}

The so-called $\ell_1$-norm $\|\beta\|_1=\sum_{j=1}^m |\beta_j|$ is
well known to promote sparsity \cite{Tibshirani1996} when used as a penalty (added to a loss
function) or as a constraint (imposed on the minimizer of a loss
function). Two major examples are the Lasso of \cite{Tibshirani1996}
(which combines a quadratic loss with an $\ell_1$-norm penalty or
constraint) and sparse quantile regression of e.g. \cite{Li2008}
(which combines the loss (\ref{lossfunction}) with an $\ell_1$-norm
penalty. See also \cite{Fuchs2009}). The advantage of using the
$\ell_1$-norm instead of the number of nonzero $\beta_j$ is
computational: the former leads to a convex minimization problem
while the latter gives rise to a combinatorial minimization problem.

In this paper our goal is to impose structure between the $m$ explanatory variables $\beta_1,\ldots\beta_m$ by dividing them into $\numgroups$ non-overlapping groups ($\numgroups\leq m$). $G_1,\ldots,G_\numgroups$ represent the indices in each group. At the same time, we desire to select a small number of active groups. We are therefore interested in imposing a penalty or constraint of $\ell_1$-norm type on the different groups. Within a given group, when a single member is active (nonzero), we allow the other members of that group to reach the same magnitude without penalty. In order to impose such a behavior within groups, the $\max$-norm is appropriate. Indeed, the value of $\max_{j\in G_k}\{|\beta_j|\}$does not increase as long as the largest of the $|\beta_j|$ (with $j\in G_k$) does not increase, regardless of the actual size of the smaller ones.

The mixed norm $\|\beta\|_{1,\infty}$ is defined as
\begin{equation}\label{mixednormdef}
\|\beta\|_{1,\infty}=\sum_{k=1}^\numgroups \|\beta_{G_k}\|_\infty,
\end{equation}
where $\beta_{G_k}$ are the components of $\beta$ in group $k$, and
$\|\beta_{G_k}\|_\infty=\max_{j\in G_k}|\beta_j|$. This mixed norm behaves as described above: Within a group, it behaves as a $\max$-norm and between groups it imposes an $\ell_1$-norm (sum of $\max$-norms).
The structured sparse $\hat{\beta}$ for the
quantile regression loss function in (\ref{loss}) is now defined as the minimizer:
\begin{equation}\label{problem}
\hat{\beta}=\arg\min_{\beta}\sum_{i=1}^{n} \varrho_{\tau}
\left((y-X\beta)_i \right) + \lambda \|\beta\|_{1,\infty}
\end{equation}
where $\lambda\geq 0$ is the penalization
parameter or as the minimizer:
\begin{equation}\label{problemR}
\hat{\beta}=\arg\min_{\|\beta\|_{1,\infty}\leq R}\,\sum_{i=1}^n \varrho_\tau\left((y-X\beta)_i\right),
\end{equation}
where $R$ is a nonnegative parameter. By suitably choosing the
parameters $\lambda$ and $R$, the minimizers of these problems are
identical. We therefore use the same symbol $\hat\beta$, both for
the minimizer of the penalized problem (\ref{problem}) and for the
minimizer of constrained problem (\ref{problemR}). We do not
indicate explicitly their dependence on $\lambda$ or $R$.

The mixed norm penalty $\|\beta\|_{1,\infty}$ has been studied
before in the framework of least squares loss functions by
\cite{Yuan2006}. A special case of that least squares problem was
already introduced by \cite{TURLACH2005}. Here, as in \cite{Zoua2008}, we combine its
structure-imposing properties with the robustness properties of
quantile regression. The aim of this paper is to present an
efficient algorithm for the solution of the minimization problems (\ref{problem}) and
(\ref{problemR}), for various values of $\lambda $ and $R$, and to illustrate its use with some applications.

\section{Solution of the minimization problem}
\label{solutionsection}

Due to the presence of the non-smooth functions $\varrho_{\tau}$ and $\|\beta\|_{1,\infty}$, the optimization problems (\ref{problem}) and (\ref{problemR}) are not differentiable. They are however convex minimization problems for which a general theory exists. We refer to e.g. \cite{Rockafellar1997} for an introduction to convex analysis. In particular, for any convex function $f$, the symbol $\partial f$ denotes the subdifferential of $f$.

We express the condition for minimizing the cost function of expression (\ref{problem}) using subdifferentials instead of usual derivatives. Necessary and sufficient conditions for optimality of the
optimization problem (\ref{problem}) are found by expressing that
$0$ belongs to the sub\-differential of the functional (\ref{problem}).
Therefore, we will have that $\beta$ is the
minimizer of (\ref{problem}) if and only if there exists a vector $w \in
\partial \sum_{i=1}^{n} \varrho_{\tau}\left((y-X\beta)_i \right)$ and a vector $u \in \partial\|\beta\|_{1,\infty}$ such that:
\begin{equation}\label{cond}
-X^T w +\lambda u=0.
\end{equation}
Setting $r=y-X\beta$, the equations that we need to solve are:
\begin{equation}\label{conditions}
\left\{
\begin{array}{l}
 -X^T w + \lambda u = 0\\[3mm]
 w_i \in \partial\varrho_{\tau}\left(r_i \right)\\[3mm]
 u \in \partial\|\beta\|_{1,\infty}\\[3mm]
 r=y-X\beta.
 \end{array}\right.
\end{equation}
Solving the minimization problems (\ref{problem}) and (\ref{problemR}) therefore requires detailed knowledge of the subdifferentials of $\varrho_{\tau}$ and of $\|\beta\|_{1,\infty}$.

In case of the function $\varrho_{\tau}$ the subdifferential is equal to:
\begin{equation}
\label{partial_easy}
\partial \varrho_{\tau}(r_i) =
\left\{
\begin{array}{rcl}
2\tau  & \text{if} & r_i>0,\\[0mm]
[-2(1-\tau),2\tau]  & \text{if} & r_i=0,\\
-2(1-\tau)& \text{if} & r_i < 0.
\end{array}
\right.
\end{equation}
It is important to remark that, conversely, the knowledge of the value of the subgradient $w_i\in \partial \varrho_{\tau}(r_i)$ also gives a certain knowledge on $r_i$. In particular, if $w_i=2\tau$, then $r_i$ must be non-negative, if $w_i=-2(1-\tau)$, then $r_i$ must be non-positive and if $w_i$ belongs to the interval $]-2(1-\tau),2\tau[$, then $r_i$ must be zero. This will be important when describing the algorithm that solves the equations (\ref{conditions}).

The subdifferential $\partial \|\beta\|_{1,\infty}$ for the penalty part $\|\beta\|_{1,\infty}$ is more difficult. As the groups are disjoint, each term in the sum (\ref{mixednormdef}) can be handled separately. Consider
group $k$ with $\|\beta_{G_k}\|_\infty=\max_{j\in G_k}|\beta_j|$,
and $G_k=\{j_1,j_2,\ldots,j_{l_k}\}$. The subdifferential of the $\max$ function is the convex hull of the union of the subdifferentials of the `maximal' arguments (see e.g. \cite{Boyd2004}). In our case, each argument of the $\max$-function is a function of just a single variable $\beta_j$. The subdifferential (w.r.t. the single variable $\beta_j$) of the absolute value $|\beta_j|$ is:
\begin{equation}
\partial |\beta_j| =
\left\{\begin{array}{rcl}
1 & \text{if} & \beta_j>0,\\[0mm]
  [-1,1]  & \text{if} & \beta_j=0,\\
-1 & \text{if} & \beta_j < 0,
\end{array}
\right.
\end{equation}
and the subdifferential of $|\beta_{\tilde\jmath}|$ w.r.t. $\beta_j$ is zero ($j\neq \tilde\jmath$).

In this way, when all components of $\beta$ in group $k$ are zero, we find that $\partial\|\beta_{G_k}\|_\infty$ consists of an $\ell_1$-ball of radius $1$.
On the other hand, when all $l_k$ coefficients in group $G_k$ have the same (nonzero) absolute size, $\partial\|\beta_{G_k}\|_\infty$ consists of a single face of an $\ell_1$-ball with radius $1$ (the signs of the $\beta_j$ determine which face). If the maximum is reached in $1,2,\ldots l_k-1$  nonzero $\beta_j$ then the set $\partial\|\beta_{G_k}\|_\infty$ (with elements $u$) consists of that part of the same face with $u_j=0$ for non-maximal components $\beta_j$ (i.e. components that are smaller than the maximum of that group:  $|\beta_j|<\|\beta_{G_k}\|_\infty$).

Here too, knowledge of $u_{G_k}\in\partial\|\beta_{G_k}\|_\infty$ yields partial information on $\beta_{G_k}$. E.g. if $\|u_{G_k}\|_1<1$ then all $|\beta_j|$ in this group are zero; if $\|u_{G_k}\|=1$ and all $u_j\neq0$ then all $\beta_j$ in that group have the same absolute value. If $u_j=0$ then $\beta_j$ may not be maximal in its group.

This type of interplay between $\beta$, $u$, $w$ and $r$ will be used frequently in the algorithm which we describe in Section~\ref{algorithmsection}. The optimality conditions (\ref{conditions}) are not analytically solvable for $\beta$. However, the problems (\ref{problem}) and (\ref{problemR}) fall into the class of problems described by \cite{ROSSET2007} which allow for a piece-wise linear solution path. This means that the minimizers of (\ref{problem}) and (\ref{problemR}) are piecewise linear functions in terms of $\lambda$ or $R$. Moreover the nodes that determine these piecewise linear functions can be calculated analytically (i.e. using linear algebra), provided one starts the calculation from $R=0$ (which corresponds to $\beta=0$ and $\lambda$ large) and proceeds carefully for increasing values of $R$ (decreasing values of $\lambda$). As explained in the next section, a new node of the minimizer $\hat\beta$ (as a function of $R$) appears e.g. when a new group becomes active (becomes nonzero) or when an equation among the $(y-X\beta)_i=0$ is satisfied. There are several more events like this possible and they are explained in more detail in the next section.

\section{Algorithm}\label{algorithmsection}

The non-iterative algorithm we propose for finding the minimizers of
(\ref{problem}) and (\ref{problemR}) is described in this section.
It is similar to the LARS algorithm of \cite{EfHJT2004} for the
Lasso and the algorithm of \cite{Li2008} and \cite{Fuchs2009} for
sparse quantile regression (where the number of groups equals the
number of explanatory variables). In our case, we are not restricted
to singleton groups; however, similar to \cite{Yuan2006}, only
non-overlapping groups are allowed.

Following \cite{Fuchs2009}, we set $u=u^{(0)}+u^{(1)}/\lambda$ and $w=w^{(0)}+\lambda w^{(1)}$, where $u^{(0)},u^{(1)},w^{(0)}$ and $w^{(1)}$ do not depend on $\lambda$. In this way equation (\ref{cond}) is equivalent to:
\begin{equation}\label{uweqns}
-X^T w^{(0)} + u^{(1)} = 0\qquad\mathrm{and}\qquad -X^T w^{(1)} + u^{(0)} = 0.
\end{equation}
As the minimizer $\hat\beta$ of the problem (\ref{problemR}) is a piecewise linear function of $R=\|\hat\beta\|_{1,\infty}$, it is characterized by a set of (interpolation) nodes. If the value of $\hat\beta$ is known in these nodes, then $\hat\beta$ can also be found for other values of $R$ by linear interpolation.

The proposed algorithm starts from $\hat\beta=0$ for $\lambda$
sufficiently large ($\lambda\geq\lambda_\mathrm{max}$). As
$\hat\beta=0$, one finds the value of $r=y-X\hat\beta=y$ and of $w$:
$w_i^{(0)}=2\tau$ if $r_i>0$, $w^{(0)}_i=-2(1-\tau)$ if $r_i<0$
(assuming that all coefficients of $r$ at this stage are nonzero)
and $w^{(1)}=0$. Using equations (\ref{uweqns}), one then calculates
$u=u^{(0)}+u^{(1)}/\lambda$. The value of $\lambda_\mathrm{max}$ is
now found by solving the equations $\|u_{G_k}\|_1=1$ (for all groups
$k:1\dots\numgroups$). This can be done on a computer as
$\|u^{(0)}+u^{(1)}/\lambda\|_{1,\infty}$ is a piecewise linear
function of $\lambda^{-1}$. The largest positive value among those
numbers is the desired value of $\lambda_\mathrm{max}$. The group
$k$ for which $\|u_{G_k}\|=1$ will be the first group to enter the
set of active groups (in the next step) and become non-zero. Once
this break point value of $\lambda$ is known, it can be used to
update $u$ and $w$.

The algorithm then continues a loop with consists of several parts:
\begin{enumerate}
\item Express $\hat\beta$ as a linear function of $R$: $\hat\beta=\hat\beta^{(0)}+R\hat\beta^{(1)}$, where $R=\|\hat\beta\|_{1,\infty}$. This uses the knowledge of the active groups, of the maximal set within each group (from the knowledge of the subgradient $u$), of the relative signs of these components of $\beta$ (also from the knowledge of $u$), and of the components of $y-X\hat\beta$ that are zero (from the knowledge of the subgradient $w$).

\item Determine the largest value of $R$ for which the expression $\hat\beta=\hat\beta^{(0)}+R\hat\beta^{(1)}$ is valid. The expression may cease to be valid when:
\begin{enumerate}
\item an additional component of $r=y-X\hat\beta$ becomes zero,
\item an active group becomes non-active (all members are zero),
\item a non-maximal component of $\hat\beta$ (in some group) becomes equal in absolute value to the maximal value in that group.
\end{enumerate}

\item Once the value of $R$ is calculated, update the variables $u$ and $w$ (as a function of $\lambda$). Here equations (\ref{uweqns}) are used together with the knowledge of $w_i$ for the nonzero $r_i$. The knowledge of the non-maximal $\beta_j$ (in each group) is also used to set some $u_j$'s to zero.

\item Calculate the smallest value of $\lambda$ for which these expressions for $u$ and $w$ are valid. The expressions for $u$ and $w$ cease to be valid when:
\begin{enumerate}
\item One of the $\|u_{G_k}\|_1$ will reach $1$ (a new active group will be added to the active set in the next step),
\item A coefficient of $w$ equals $2\tau$ or $-2(1-\tau)$ (in this case, an equation $r_j=0$ that is satisfied in the current step, will no longer be satisfied in the next step),
\item in an active group $k$, one of the coefficients of $u_{G_k}$ becomes equal to $0$. In this case, the corresponding component of $\hat\beta$ will be of smaller absolute value than the maximal value in that group in the next step.
\end{enumerate}

\item Continue with step 1 or stop.
\end{enumerate}
The algorithm may be stopped when the desired maximum number of active groups is reached, when $\lambda=0$, or when some other suitable stopping criteria is satisfied.

In this algorithm, steps 1 and 3 require the solution of a linear system of equations.
Steps 2 and 4 require the solution of simple linear equations to determine $R$ or $\lambda$ at break points. Here numerical round-off error may affect the accuracy of these calculations. Unfortunately, the decisions (groups entering or leaving the active set, coefficients becoming submaximal in a group, \ldots) depend on these numerical results. Round-off errors may therefore lead to the wrong decisions being taken by the algorithm. In that case the algorithm fails. This shortcoming is common to all the algorithms of this type \cite{EfHJT2004,Li2008,Fuchs2009}.

When dealing with data $X$ and $y$ containing small integers, or when rows or columns of $y$ and $X$ repeat, it is possible that different events (2a--c or 4a--c) occur simultaneously. One could e.g. have two groups enter the active set at the same step. Another possibility is that a new group becomes active at the same step when a component of an active group becoming sub-maximal. Such possibilities are not accounted for in the current implementation of the algorithm.  This ``one-at-a-time condition'' \cite[p417]{EfHJT2004} is also common to algorithms of this type (the work of \cite{EfHJT2004,Li2008,Fuchs2009} also does not handle such cases).
\cite[p438]{EfHJT2004} have proposed to add a small amount of jitter to the variables to overcome the problem.

The most effective way of understanding the proposed algorithm is by going through a worked-out example step-by-step. Table~\ref{exampletable} lists the complete solution path of a simple example with:
\begin{equation}\label{exampleX}
X=
\left(\begin{array}{ccc}
-4 & 3 & 5\\
-4 & 5 & 1\\
4 & -3 & 0
\end{array}\right)
,\qquad y=
\left(\begin{array}{c}
8\\
7\\
-11
\end{array}\right),
\end{equation}
groups $G_1=\{1\}$ and $G_2=\{2, 3\}$ and $\tau=1/2$. The solution
path is given as a function of $R$ and $\lambda$, and the
intermediate values of the subgradients $u$ and $w$ are also given.
This example was chosen in such a way that every possibility in
steps 2 and 4 of the algorithm occurs at least once.

\begin{figure}\centering
\resizebox{\textwidth}{!}{\includegraphics{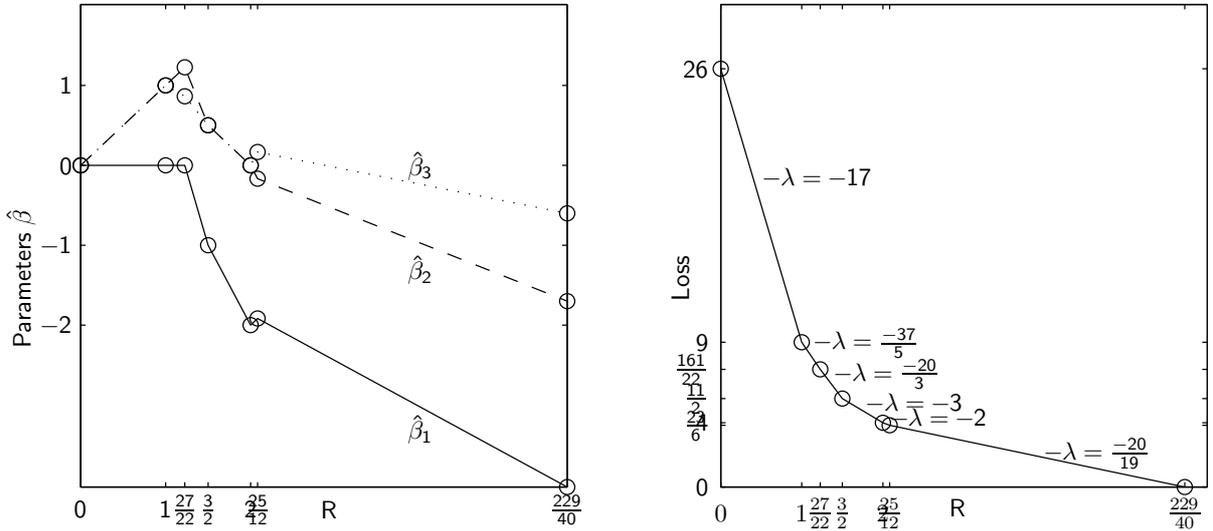}}
\caption{Minimizers corresponding to the example $X$ and $y$ of equation (\ref{exampleX}). Left: parameters $\hat\beta$ as a function of $R=\|\hat\beta\|_{1,\infty}$. Right: the corresponding trade-off curve plotting the loss $\sum_{i=1}^n\varrho_\tau((y-X\hat\beta)_i)$ as a function of the  penalty $R=\|\hat\beta\|_{1,\infty}$. The slope of this curve is equal to $-\lambda$.}\label{tradeoffpic}
\end{figure}

As one can see in the example, there are certain values of $\lambda$ (i.e. $\lambda=17,37/5,20/3, \ldots$) for which the minimizer of the penalized problem (\ref{problem}) is not unique. One also sees that between these special values of $\lambda$ the minimizer $\hat\beta$ of (\ref{problem}) is constant as a function of $\lambda$ (the subgradients $u$ and $w$ do change). This behavior is easy to interpret by plotting the loss $\sum_{i=1}^n\varrho_\tau((y-X\hat\beta)_i)$ as a function of $R=\|\hat\beta\|_{1,\infty}$, as was done for example (\ref{exampleX}) in Figure~\ref{tradeoffpic}. We see that the graph is piecewise linear and that
$\lambda$ is locally equal to the slope of this trade-off curve.
Therefore, between break points, $\lambda$ is constant (several $\hat\beta$ correspond to the same value of $\lambda$) and at break points, $\lambda$ takes on several values (in other words, for several values of $\lambda$, the solution $\hat\beta$ of (\ref{problem}) is constant).

A set of Matlab functions that implement the above algorithm was written by the authors, and is available on their web page \cite{Lorisweb2012}.

\begin{table}\centering%
\resizebox{\textwidth}{!}{%
\begin{tabular}{lccccccl}
\# & $\hat\beta$ & $r$ & $R$ & $\lambda$ & $u$ & $w$ & Step\\ \hline
\rule{0mm}{5mm}0& $(0, 0, 0)$ & $(8, 7, -11)$ & $0$ & $ [17,+\infty[ $ &$\lambda^{-1}(-12, 11, 6)$ &$(1, 1, -1)$ & 
\\
& $(0,R,R)$& $$\vdots$$ & $[0,1]$ & $17$ &$(\frac{-12}{17}, \frac{11}{17}, \frac{6}{17})$ & $(1, 1, -1)$& 1\,\raisebox{-3.5mm}[0pt][0pt]{$\downarrow$2a} \\
1& $(0, 1, 1)$ & $(0, 1, -8)$ & $1$ & $ [\frac{37}{5},17] $ &$\vdots$ &$\vdots$ & 3\,\raisebox{-3.5mm}[0pt][0pt]{$\downarrow$4c}\\
& $\vdots$& $$\vdots$$ & $[1,\frac{27}{22}]$ & $\frac{37}{5}$ &$(\frac{-36}{37}, 1, 0)$ & $(\frac{-1}{5}, 1, -1)$& 1\,\raisebox{-3.5mm}[0pt][0pt]{$\downarrow$2a} \\
2& $(0, \frac{27}{22}, \frac{19}{22})$ & $(0, 0, \frac{-161}{22})$ & $\frac{27}{22}$ & $ [\frac{20}{3},\frac{37}{5}] $ &$\vdots$ &$\vdots$ & 3\,\raisebox{-3.5mm}[0pt][0pt]{$\downarrow$4a}\\
& $\vdots$& $$\vdots$$ & $[\frac{27}{22},\frac{3}{2}]$ & $\frac{20}{3}$ &$(-1, 1, 0)$ & $(\frac{-1}{6}, \frac{5}{6}, -1)$& 1\,\raisebox{-3.5mm}[0pt][0pt]{$\downarrow$2c} \\
3& $(-1, \frac{1}{2}, \frac{1}{2})$ & $(0, 0, \frac{-11}{2})$ & $\frac{3}{2}$ & $ [3,\frac{20}{3}] $ &$\vdots$ &$\vdots$ & 3\,\raisebox{-3.5mm}[0pt][0pt]{$\downarrow$4b}\\
& $\vdots$& $$\vdots$$ & $[\frac{3}{2},2]$ & $3$ &$(-1, \frac{1}{12}, \frac{11}{12})$ & $(\frac{3}{4}, -1, -1)$& 1\,\raisebox{-3.5mm}[0pt][0pt]{$\downarrow$2b} \\
4& $(-2, 0, 0)$ & $(0, -1, -3)$ & $2$ & $ [2,3] $ &$\vdots$ &$\vdots$ & 3\,\raisebox{-3.5mm}[0pt][0pt]{$\downarrow$4a}\\
& $\vdots$& $$\vdots$$ & $[2,\frac{25}{12}]$ & $2$ &$(-1, \frac{-1}{4}, \frac{3}{4})$ & $(\frac{1}{2}, -1, -1)$& 1\,\raisebox{-3.5mm}[0pt][0pt]{$\downarrow$2a} \\
5& $(\frac{-23}{12}, \frac{-1}{6}, \frac{1}{6})$ & $(0, 0, \frac{-23}{6})$ & $\frac{25}{12}$ & $ [\frac{20}{19},2] $ &$\vdots$ &$\vdots$ & 3\,\raisebox{-3.5mm}[0pt][0pt]{$\downarrow$4c}\\
& $\vdots$& $$\vdots$$ & $[\frac{25}{12},\frac{229}{40}]$ & $\frac{20}{19}$ &$(-1, -1, 0)$ & $(\frac{7}{38}, \frac{-35}{38}, -1)$& 1\,\raisebox{-3.5mm}[0pt][0pt]{$\downarrow$2a} \\[2mm]
6& $(\frac{-161}{40}, \frac{-17}{10}, \frac{-3}{5})$ & $(0, 0, 0)$ & $\frac{229}{40}$ & $ [0,\frac{20}{19}] $ &$(-1, -1, 0)$ &$\lambda(\frac{7}{40}, \frac{-7}{8}, \frac{-19}{20})$ & 3\,\raisebox{-3.5mm}[0pt][0pt]{\rule{2mm}{0mm}}\\
\end{tabular}
}
\caption{The various values of $\hat\beta, r, R, \lambda, u, w$ etc. that describe the minimizers (\ref{problem}) and (\ref{problemR}) for the simple example (\ref{exampleX}). The last column indicates the step that is taken in the algorithm of Section~\ref{algorithmsection}. Values of $\hat\beta$, $r$, $u$ or $w$ that can be found through interpolation (w.r.t $R$ for $\beta$ and $r$, w.r.t $\lambda$ for $w$ and w.r.t. $\lambda^{-1}$ for $u$) have been replaced by vertical dots for lack of space.}\label{exampletable}
\end{table}

\section{Applications}\label{applicationsection}

The algorithm presented in Section~\ref{algorithmsection} gives the
entire solution path for different values of $R=\|\hat\beta\|_{1,\infty}$ in some interval
$[0, R_\mathrm{max}]$. Selecting the appropriate value of $R$ (and
the corresponding coefficients $\hat\beta$) is an important issue
for practical purposes. As \cite{Koenke1994} have proposed, the Bayesian
information criterion (BIC) of \cite{Schwarz1978} is a promising
information criterion for model selection in quantile regression.
In view of Theorem~2 of \cite{Li2008}, and the
loss function in (\ref{lossfunction}), the adapted BIC for quantile
regression is as follows:
\begin{equation}\label{bic}
    \mbox{BIC}(R)=\log\left( \frac{1}{n}\sum_{i=1}^n \varrho_\tau\left((y-X\hat{\beta})_i\right)\right)-\frac{\log (n)}{2n}\,\,\,n_R,
\end{equation}
where $\hat\beta$ is a function of $R$ and $n_R$ is defined as the
number of zeros in the residual vector $r=y-X\hat\beta$. The model with
smaller BIC is more desirable (see e.g. \cite{Hjort2008}).

In this section two main applications of structured
sparse quantile regression are studied using real data sets as illustrations.

\subsection{Low birth weight data set}\label{lbwsection}

According to \cite{Xyz}, \emph{low birth weight} (LBW) is defined as
a birth weight of a liveborn infant of less than $2500$g regardless
of gestational age. LBW has negative effect both on the infants and
the parents, e.g mothers of LBW babies have a greater chance of
having postpartum depression and they need more time before
returning to work, infants who are born with LBW are at greater risk
of having learning or vision difficulties. Also, LBW infants would
impose large costs on society. The risks of LBW are discussed by
many authors such as, \cite{Boardman2002}, \cite{Auslander2003} and
\cite{Almond2005}. Therefore, determining the effective
factors in LBW infants is very important. If one is interested in studying the effects of different factors on the
lower tail of the conditional distribution of infants' weight,
\cite{Koenker2001} have remarked that using the least squares
regression methods (e.g. \cite{Yuan2006}) is not reasonable. Using
quantile regression with $\tau$-th quantile ($\tau\leq 0.5$) would
give the possibility to study lower tail of the infants weight
given the explanatory variables.

As in \cite{Yuan2006}, the data are taken from \cite{Lameshow1989}.
The data set contains the birth weight (expressed in grams) of $189$
infants as the response variable and $8$ explanatory variables:
mother's age (in years), mother's weight (in pounds), mother's race
($(1,0)=$ black, $(0,1)=$ white or $(0,0)=$ other), smoking status
during pregnancy ($1=$yes or $0=$no), number of previous premature
labours ($0, 1, 2, \ldots$), history of hypertension ($1=$ yes or
$0=$ no), presence of uterine irritability ($1=$ yes or $0=$ no),
number of physician visits during the first trimester ($0, 1, 2, \ldots$). The data were collected at Baystate Medical Center,
Springfield, Massachusetts, in the year 1986.

As \cite{Koenker2001} and \cite{Yuan2006} have suggested, some
non-linear effects of two of the quantitative predictors (mother's
weight and age) may exist. Therefore, in accordance with
\cite{Koenker2001} we consider a second-order polynomial for both of
them. We put each corresponding pair of variables in one group.
A variable which needs a non-singleton group is race (it is a
nominal variable with more than two levels). Thus in order to study
its effect one may create two dummy variables out of it (see also Section~\ref{introductionsection}). As was already mentioned
in the introduction, both or none of them should be included in the
model. So we may put them both in one group. All other groups are
singletons.

As it is mentioned in \cite{Koenker2001}, and considering Tukey's
dictum: \emph{Never estimate intercepts, always estimate
centercepts}, the quantitative variables in the model are centered
and re-scaled by dividing by their standard deviations. Therefore,
the estimated intercept may be interpreted as the weight of an
infant born to a $23$ year old mother, whose weight was $130$
pounds, her race was other (neither black nor white), she was a
non-smoker during her pregnancy, with on average $0.16$  previous premature labours, no history of hypertension,
no presence of irritability, and an average of $0.47$ physician visits during the
first trimester.

A small amount of jitter is added to the variables $y$ and $X$, to guarantee that the one-at-a-time condition mentioned at the end of Section~\ref{algorithmsection} is satisfied (see e.g. also
\cite[p438]{EfHJT2004}). We have verified that this does not change the outcome of the numerical experiments.

\begin{figure}
\resizebox{\textwidth}{!}{\includegraphics{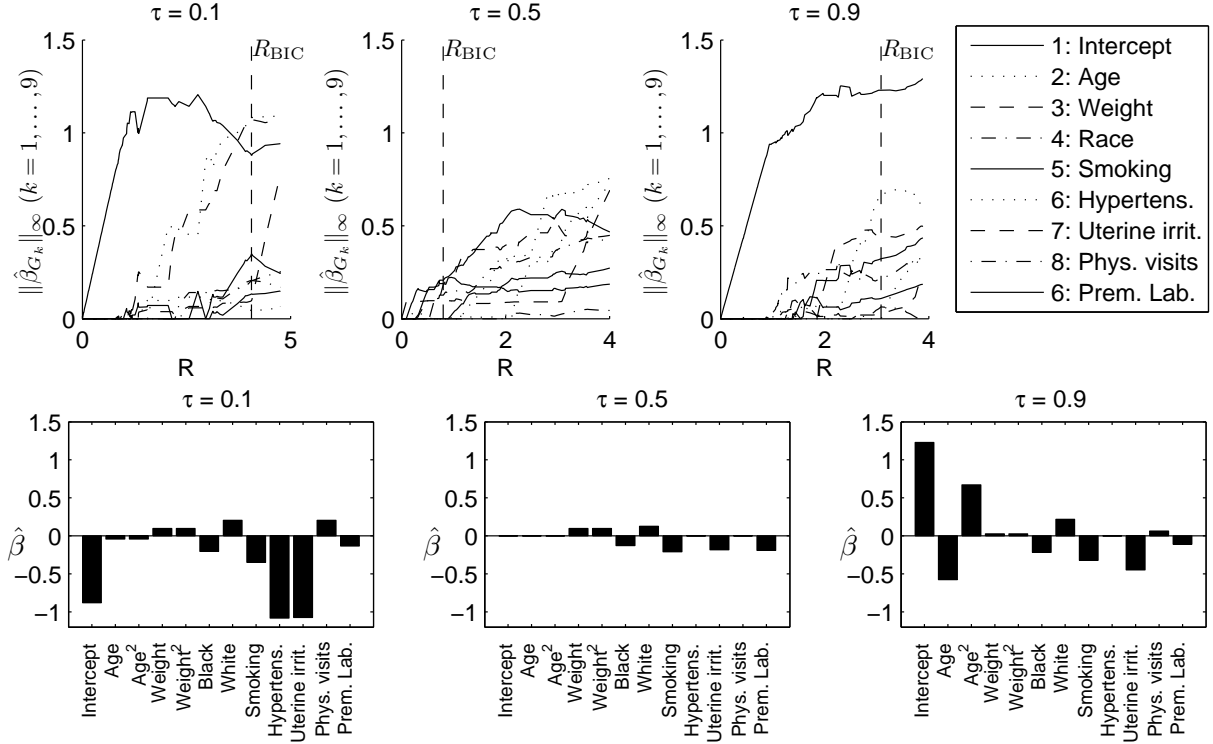}}
\caption{Analysis of the LBW data set. The first row shows (part of) the solution path given by the algorithm in Section~\ref{algorithmsection} for $\tau=0.1,
0.5, 0.9$. The vertical dashed line indicates the model chosen by the BIC (for $R=R_\mathrm{BIC}$). The second row shows the corresponding coefficients $\hat{\beta}$ for the BIC model.}\label{path}
\end{figure}

Figure~\ref{path} (top) presents (part of) the solution path of the
model coefficients $\hat\beta$ for for $\tau=0.1$,
$\tau=0.5$, and $\tau=0.9$. A vertical dashed
line is drawn at $R=R_\mathrm{BIC}$ to indicate the coefficients chosen by
the BIC (\ref{bic}). The bottom row of this figure contains the
three corresponding model coefficients (chosen by BIC). As one may
see, the models chosen for various values of $\tau$ are different,
i.e. the effective variables for the lower tail of the conditional distribution are different from the ones for the median and for the upper tail. Since we are interested in
studying the effective variables on the LBW, using loss functions
such as least-squares (with minimizer equal to the conditional
mean) would only give partial information. This result is in accordance with
\cite{Koenker1999} who remarked that the effects of the explanatory
variables in LBW data set are not constant across the conditional
distribution of the infant's weight. In other words, for different
quantiles one may have different models.

\subsection{Simultaneous variable selection for a vector of response variables}

Consider $Y=(y_1,y_2,\ldots,y_p)$ a vector of possibly correlated
response variables. One is interested in modelling these variables using
a common set of explanatory variables $X=(x_1,x_2,\ldots,x_m)$. For
variable selection purposes, one may consider $p$ linear models
$y_j=X\beta+\epsilon$, ($j=1,\ldots,p$) separately and find e.g. the
penalized model solution for $\beta$ for each model. But as
\cite{TURLACH2005} have suggested, it is sometimes interesting to
select variables by considering all $p$ response variables
simultaneously, specially when these $p$ variables are correlated.

Suppose we have $n$ observations $y_{ij}$ (with $i=1,\ldots, n$ and
$j=1,\ldots, p$) for $Y=(y_1,y_2,\ldots,y_p)$ as response variables
and $X=(x_1,x_2,\ldots,x_m)$ as explanatory variables. Let
$\beta_{kj}$, ($k=1,\ldots,m$ and $j=1,\ldots,p$) be the regression
coefficient of $x_k$ regressed on $y_j$. As \cite{TURLACH2005} have
observed,
$\|(\beta_{k1},\beta_{k2},\ldots,\beta_{kp})\|_\infty
=\max\{|\beta_{k1}|,|\beta_{k2}|,\ldots,|\beta_{kp}|\}$
is a reasonable measure of the explanatory power of the regressor
$x_k$ on all $p$ response variables $y_j$ simultaneously. If the
least-squares loss of \cite{TURLACH2005} is replaced by the more
robust loss function (\ref{loss}), the following optimization
problem:
\begin{equation}\label{problem_application}
\hat\beta=\arg\min_{\beta} \sum_{i=1}^n \sum_{j=1}^p
\varrho_{\tau}\left(y_{ij}-\sum_{k=1}^m x_{ik}\beta_{kj}\right)
+ \lambda \sum_{k=1}^m \max\{|\beta_{k1}|, |\beta_{k2}|, \ldots, |\beta_{kp}|  \}.
\end{equation}
should be solved to select the variables. Equivalently, one could also use the constrained formulation taking the form:
\begin{equation}\label{problem_application2}
\hat\beta=\arg\min_{\sum_{k=1}^m \max\{|\beta_{k1}|, |\beta_{k1}|, \ldots, |\beta_{kp}|  \}\leq R} \sum_{i=1}^n \sum_{j=1}^p
\varrho_{\tau}\left(y_{ij}-\sum_{k=1}^m x_{ik}\beta_{kj}\right).
\end{equation}
As the argument of $\varrho_\tau$ in these last two expressions
is a linear function of the $\beta_{kj}$, problems
(\ref{problem_application}) and (\ref{problem_application2}) are
special cases of problems (\ref{problem}) and (\ref{problemR}), and
can therefore be solved by the algorithm of
Section~\ref{algorithmsection}. As \cite{TURLACH2005} pointed out,
this is an exploratory tool for identifying a suitable subset of
regressor variables, not for actual parameter estimation. The
problem (\ref{problem_application}) has already been proposed by
\cite{Zoua2008} who solved it for a fixed value of $R$ using a
generic linear programming code. The algorithm of
Section~\ref{algorithmsection} finds the minimizer for a whole range
of values of $R$. The latter algorithm is therefore more useful as
the BIC criterion (\ref{bic}) (also used by \cite{Zoua2008})
requires a further minimization over many values of the parameter
$R$.

As an example, we consider the `93CARS' data set which contains
information on $93$ new cars for the 1993 model year which is
obtained from \cite{Amstat}. Table \ref{Variables} presents the
variables we have considered. Some of the observations have been
omitted due to missing values, so in total $82$ observations are
used.

\begin{table}[h]\centering
\caption{Variables in 93CARS data set.}\label{Variables}
\begin{tabular}{rll}
&Variable & Description \\ \hline
$y_1$:& Minimum price (in \$1,000) & Price for basic version of this model \\
$y_2$:& Midrange price (in \$1,000) & Average of Min and Max prices\\
$y_3$:& Maximum price (in \$1,000) & Price for a premium version\\
$y_4$:& City MPG & miles per gallon by EPA rating \\
$y_5$:& Highway MPG &  -\\
$x_1$:& Number of cylinders& -\\
$x_2$:& Engine size & in liters \\
$x_3$:& Horsepower   & maximum\\
$x_4$:& RPM & revs per minute at maximum horsepower\\
$x_5$:& Engine revolutions per mile& (in highest gear)\\
$x_6$:& Fuel tank capacity & in gallons \\
$x_7$:& Passenger capacity & in persons \\
$x_8$:& Length & in inches \\
$x_9$:& Wheelbase & in inches \\
$x_{10}$:& Width & in inches\\
$x_{11}$:& U-turn space & in feet\\
$x_{12}$:& Rear seat room & in inches \\
$x_{13}$:& Luggage capacity & in cubic feet\\
$x_{14}$:& Weight & in pounds
\end{tabular}
\end{table}

\begin{figure}
\resizebox{\textwidth}{!}{\includegraphics{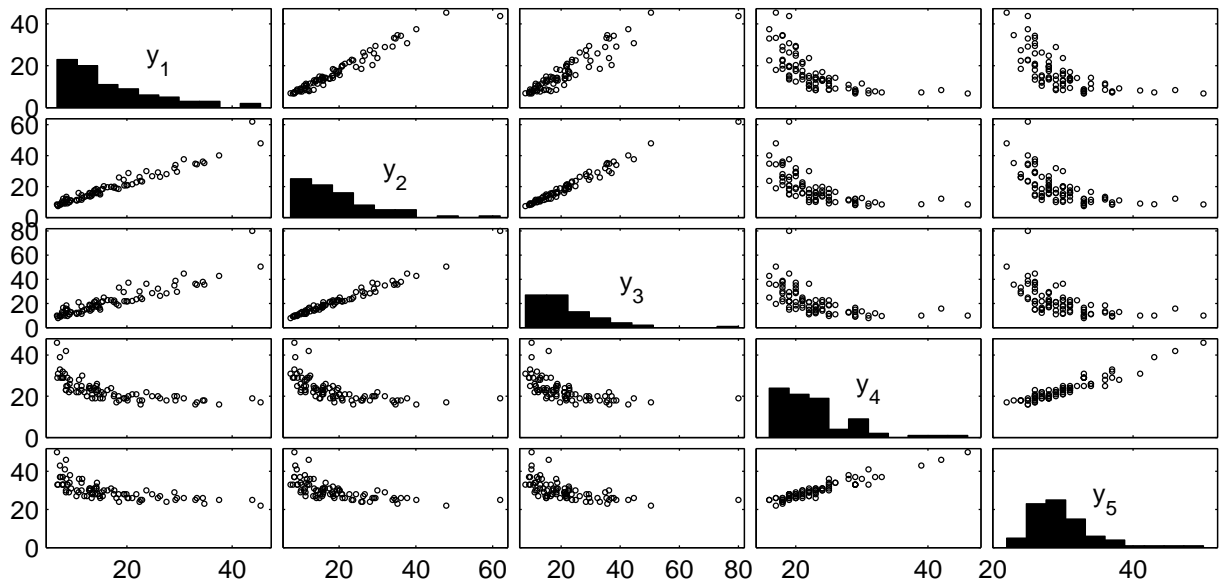}}
\caption{Pairwise scatter plots of the $5$ response variables in the
93CARS data set. The histogram of each response variable $y$
is presented on the diagonal.}\label{scatter}
\end{figure}

Figure \ref{scatter} shows the pairwise scatter plots of the $5$ response
variables. The two main points which follow from this figure
are: 1. the variables are correlated, and 2. there are some outliers in the data. Therefore, a simultaneous variables selection using
least absolute deviation (QR with $\tau=0.5$) seems reasonable here.
Both response variables $y$ and regression variables $X$ are standardized, so one would be
able to compare variables in different measures.

\begin{figure}
\resizebox{\textwidth}{!}{\includegraphics{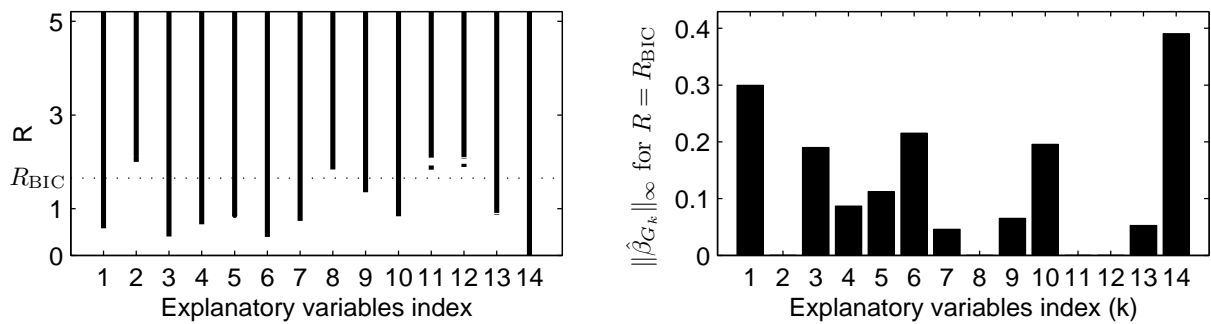}}
\caption{Structured sparse quantile regression of the 93CARS data set. Left: the $\ell_{\infty}$-norm of coefficients
$\hat{\beta}$'s in each group (vertical) as a function of $R$
(horizontal). Right: (top) presence or absence of each group along
the solution path, (bottom) the $\ell_{\infty}$-norm of coefficients
$\hat{\beta}$ in each group for $R=R_\mathrm{BIC}$ (selected using the BIC (\ref{bic})).}\label{path2}
\end{figure}

The algorithm of Section~\ref{algorithmsection} is used with
$\tau=0.5$. In this application too, a small amount of jitter is
added to the variables $y$, so as to guarantee that the
one-at-a-time condition mentioned at the end of
section~\ref{algorithmsection} is satisfied (see e.g. also
\cite[p438]{EfHJT2004}). We have again verified that this does not
change the outcome of the numerical experiments.
Figure \ref{path2} (left) shows the presence or
absence of each group (as a functions of $R=\|\beta\|_{1,\infty}$).

The BIC (\ref{bic}) is used to select a model among all possible
models calculated along the solution path. The corresponding value
of $\|\hat\beta\|_{1,\infty}$ is called $R_\mathrm{BIC}$, and is indicated
on the first panel with a dotted line. The final model is given in
Figure~\ref{path2}, (right). As one
may see, the variables $x_2, x_8, x_{11}$ and $x_{12}$ were not
selected.

\section{Conclusions}\label{conclusionsection}

A structured sparse solution (or group sparse solution) of a quantile regression model,
based on penalizing or constraining the quantile regression loss function by a mixed $\ell_{1,\infty}$-norm of regression coefficients, was discussed.

An algorithm to compute the solution of the corresponding minimization problem was presented. This algorithm computes the minimizer of the penalized or constrained loss function for all values of $R=\|\hat\beta\|_{1,\infty}$ within a range $[0,R_\mathrm{max}]$ instead of just for a single value of $R$. This is a strong point when using the BIC criterion (which needs a further minimization over $R$) for model selection.

In a first application, the effective variables for the lower and upper quantiles of the conditional distribution of the birth weight of infants in the LBW data set were identified, subject to a group sparsity constraint. As a second application we studied the problem of simultaneous variable selection in a quantile regression model for robustly modeling a vector of possibly correlated response variables using a common set of explanatory variables.

The implementation of the algorithm presented in Section~\ref{algorithmsection} is not straightforward. Therefore, such an implementation in Matlab is provided on the authors' webpage
\cite{Lorisweb2012}, together with the scripts for processing the LBW data set and the 93CARS data set. The necessary functions for interpolating the solution $\hat{\beta}$ between the nodes and choosing the best model using the proposed BIC (\ref{bic}) are also provided.

The current short article dealt with the non-overlapping group case. An potential extension of the algorithm would consist of including the overlapping group case as well.

\section*{Acknowledgements}
I. L. is a research associate of the F.R.S.-FNRS  (Belgium). This
research was  supported by VUB GOA-062 and by the FWO-Vlaanderen
grant G.0564.09N.

\bibliography{groupqr}{}
\bibliographystyle{plain}

\end{document}